\documentclass{svjour2}                    
\smartqed  
\usepackage{mathptmx}      
\usepackage[numbers]{natbib}
\usepackage{graphicx}
\usepackage{bm}

\DeclareMathAlphabet{\bs}{OT1}{ptm}{b}{it}
\newcommand{\eo}{\bs{e}_0}
\newcommand{\vd}{v_{\rm d}}            
\newcommand{\vs}{v_{\rm s}}            
\newcommand{\DR}{D_{\rm R}}            
\newcommand{\zf}{z_{\rm f}}            
\newcommand{\tf}{{t_{\rm f}}}          
\newcommand{\ap}{\alpha_{\rm p}}       
\newcommand{\zhat}{\hat{\bs{z}}}       
\newcommand{\nablac}{|\nabla c|}       

\journalname{Bulletin of Mathematical Biology}
\begin{document}

\title{Run and tumble chemotaxis in a shear
flow: the effect of temporal comparisons and other complications
\thanks{J. T. Locsei is supported by an Oliver Gatty Studentship from
the University of Cambridge.}
}


\author{J. T. Locsei         \and
        T. J. Pedley 
}


\institute{J. T. Locsei \at
              Department of Applied Mathematics and Theoretical Physics,
              University of Cambridge, Cambridge CB3 9EW, U.K. \\
              \email{j.t.locsei@damtp.cam.ac.uk}           
           \and
           T. J. Pedley \at
              Department of Applied Mathematics and Theoretical Physics,
              University of Cambridge, Cambridge CB3 9EW, U.K.
}

\date{Received: date / Accepted: date}

\maketitle

\begin{abstract}
\emph{Escherichia coli} is a motile bacterium that moves up a
chemoattractant gradient by performing a biased random walk composed of
alternating runs and tumbles. This paper presents calculations of the
chemotactic drift velocity $\vd$ (the mean velocity up the
chemoattractant gradient) of an \emph{E. coli} cell performing
chemotaxis in a uniform, steady shear flow, with a weak chemoattractant
gradient at right angles to the flow. Extending earlier models, a
combined analytic and numerical approach is used to assess the effect
of several complications, namely (i) a cell cannot detect a
chemoattractant gradient directly but rather makes temporal comparisons
of chemoattractant concentration, (ii) the tumbles exhibit persistence
of direction, meaning that the swimming directions before and after a
tumble are correlated, (iii) the cell suffers random re-orientations
due to rotational Brownian motion, and (iv) the non-spherical shape of
the cell affects the way that it is rotated by the shear flow. These
complications influence the dependence of $\vd$ on the shear rate
$\gamma$. When they are all included, it is found that (a) shear
disrupts chemotaxis and shear rates beyond $\gamma \approx 2 {\rm
s}^{-1}$ cause the cell to swim down the chemoattractant gradient
rather than up it, (b) in terms of maximising drift velocity,
persistence of direction is advantageous in a quiescent fluid but
disadvantageous in a shear flow, and (c) a more elongated body shape is
advantageous in performing chemotaxis in a strong shear flow.

\keywords{Chemotaxis \and Shear flow \and Random walk \and Escherichia
coli} \subclass{62P10 \and 82B41 \and 92B05 \and 76Z10}
\end{abstract}

\section{Introduction}
\label{sec:introduction}

Free swimming bacteria are present in many naturally occurring aqueous
environments including animal intestines (enteric bacteria) and the
open ocean (marine bacteria). In the ocean, motility could influence
bacterial ecology and the role of bacteria in oceanic biogeochemistry
\citep{Grossart.2001}. Thus there is an oceanographic motivation to
understand how motile bacteria behave in a weakly turbulent, sheared
environment. Bacteria are also known to adhere to each other and/or
other surfaces and form biofilms \citep{Costerton.1995}. To understand
the formation of biofilms in the presence of a fluid flow it may be
helpful to model the motion of free swimming bacteria in the sheared
region adjacent to the biofilm.

In this paper we focus on modelling \emph{Escherichia coli}, which is
the most well studied motile bacterium. As described by
\citet{Berg.1983}, \emph{E. coli} is a common enteric bacterium with a
rod shaped body $\approx 1\;\mu{\rm m}$ in diameter and $\approx
2\;\mu{\rm m}$ long. A typical \emph{E. coli} cell has $\approx 6$
left-handed helical flagella emerging from random points on the sides
of its body, each of them extending several body-lengths into the
surrounding fluid. The flagella are powered by reversible rotary
motors. When all the flagella spin counter-clockwise, they form a
synchronous bundle and propel the cell forward in an almost-straight
`run'. When one or more flagella spin clockwise, the flagellar bundle
comes apart and the cell swims in a highly erratic `tumble', with
little net displacement but a large change in direction. Tumbles
exhibit `persistence of direction', meaning that the swimming
directions before and after a tumble are correlated, with a mean angle
between the new and previous swimming directions of $\approx 62^{\rm
o}$ \citep{Berg.1972}. Hereafter, for brevity,  `persistence of
direction' shall be referred to simply as `persistence'.

Even during a run, the cell exhibits small, random changes in
direction, and these have been attributed to rotational Brownian motion
(\emph{i.e.} thermal collisions with molecules in the surrounding
fluid). \citet{Berg.1983} estimated the coefficient of Brownian
rotation to be $\DR \approx 0.062\;{\rm radians}^2{\rm s}^{-1}$ for an
\emph{E. coli} cell swimming in a fluid of viscosity 2.7 cp at $32^{\rm
o}{\rm C}$, and this is consistent with experimental observations
\citep{Berg.1972}. However, this estimate was based on treating the
\emph{E. coli} cell as a sphere of diameter $2\;{\rm \mu m}$, and when
one takes into account the stabilising effect of the flagellar bundle,
the theoretically predicted coefficient of Brownian rotation is an
order of magnitude smaller than the observed rotational diffusivity
\citep{HenryFuCommunication}, so it seems likely that the observed
diffusivity is in fact due to intrinsic `wobbly swimming' rather than
true (thermal) Brownian rotation \citep{HowardBergCommunication}.

An \emph{E. coli} cell performs chemotaxis (\emph{i.e.} swims up a
chemoattractant gradient) by alternately running and tumbling and
biasing the length of the runs. The run durations are exponentially
distributed with a rate constant (`tumble rate') of $\lambda \approx 1
\; {\rm s^{-1}}$, and a corresponding mean run duration of
$\lambda^{-1} \approx 1 \; {\rm s}$ \citep{Berg.1972}. Tumble durations
are also exponentially distributed but with a much shorter mean
duration of $\approx 0.1 \; {\rm s}$. An \emph{E. coli} cell is too
small to detect spatial differences in the concentration of a
chemoattractant on the scale of the cell length, so it performs
temporal comparisons instead. It continually `measures' the
concentration of chemoattractants (typically nutrients) in its
environment such as serine and aspartate. If the concentration over the
past second is higher than the concentration over the previous three
seconds, the tumble rate is reduced and the expected run length is
extended. Thus, the cell performs a biased random walk and gradually
drifts toward regions of high chemoattractant concentration. The cell's
mean velocity up the chemoattractant gradient is termed the `drift
velocity'.

There have been a number of theoretical studies on run and tumble
chemotaxis with no background flow \citep{Schnitzer.1993,
deGennes.2004, Erban.2004, Erban.2005, Clark.2005, Locsei.2007}, and on
the role of bacterial shape in chemotaxis \citep[][and references
therein]{Young.2006}. Some theoretical work has also been published on
run and tumble chemotaxis with background flows. \citet{Bowen.1993}
simulated bacterial chemotaxis toward a neutrally buoyant phytoplankton
cell exuding dissolved organic carbon in an unsteady shear flow.
\citet{Luchsinger.1999} simulated a similar situation, but also
investigated the case in which bacteria reverse direction rather than
tumble. \citet{Bearon.2000} and \citet{Bearon.2003} developed an
advection-diffusion equation describing run and tumble chemotaxis in a
shear flow. The work of \citet{Bearon.2000} provided a particular
motivation for the present study, since it showed that time-delays in
the bacterial response can have a strong effect on drift velocity in a
shear flow.

The present work is in one sense an extension of \citet{Locsei.2007} to
include shear flows, and in another sense an extension of
\citet{Bearon.2000} to include temporal comparisons, non-spherical cell
shape, persistence, and Brownian rotation. Section \ref{sec:model}
introduces the model and its basic assumptions, and section
\ref{sec:choosingR} describes in more detail the response function that
is used to model the temporal comparisons performed by the cell.
Section \ref{sec:vd} presents a general analytic framework for
calculating the drift velocity. The remainder of the paper is devoted
to specific calculations of the drift velocity while taking account of
(i) temporal comparisons performed by the cell, (ii) persistence of
direction, (iii) Brownian rotation, and (iv) non-spherical cell shape.
The drift velocity under the combined effects (i), (ii) and (iii) is
computed analytically in section \ref{sec:vdspherical}, and the drift
velocity under the combined effects (i) and (iv) is computed
semi-analytically in section \ref{sec:vdspheroid}. The case where all
effects (i)--(iv) are present is treated numerically by Monte Carlo
simulation in section \ref{sec:vdsimulation}. Section
\ref{sec:conclusion} summarises the key findings.

\section{Outline of model}
\label{sec:model}

Consider a cell performing run and tumble chemotaxis, swimming in an
unbounded fluid with a chemoattractant concentration gradient $\nabla c
\parallel \zhat$ and with a background shear flow $\bs{u} =
\gamma \,z \, \hat{\bs{x}}$ where $\hat{\bs{x}}$ and $\zhat$ are unit
vectors in the $x$ and $z$ directions and $\gamma$ is the shear rate.
For convenience, we shall frequently parameterise the strength of the
shear flow by $\Omega = \gamma/2$ rather than $\gamma$. We assume that
during a run the cell swims at constant speed $\vs$ relative to the
fluid. The cell's swimming direction is described by a unit vector
$\bs{e}$ that changes with time due to tumbles, Brownian rotation, and
the shear flow.

During a run, the probability that the cell tumbles in the next time
interval $dt$ is $\lambda(t) dt$, where $\lambda$ is the `tumble rate'.
We assume that the cell alters its tumble rate linearly in response the
chemoattractant, according to
\begin{equation}
\label{eq:lambda}
   \lambda(t)=\lambda_0
      \left[
         1 - \Delta(t)
      \right],
\end{equation}
where $\lambda_0 = 1 {\rm s}^{-1}$ is the baseline tumble rate, and the
fractional change in tumble rate is given by
\begin{equation}
\label{eq:Delta}
  \Delta(t)=\int_{-\infty}^t c(t') R(t-t') dt',
\end{equation}
where $c(t')$ is the chemoattractant concentration experienced by the
cell at time $t'$, and $R$ is the cell's `response function'. $R$ may
be thought of as the impulse response of the tumble rate, since it
describes the way that $\lambda$ changes when the cell is subject to a
delta function impulse of chemoattractant concentration. The form of
tumble modulation given by (\ref{eq:Delta}) has been used in several
earlier models \citep{Schnitzer.1993, deGennes.2004, Clark.2005,
Locsei.2007} and it is motivated by the experimental results of
\citep{Block.1982} and \citep{Segall.1982}. The form of $R$ is
discussed in section \ref{sec:choosingR}. Our analysis will be
restricted to `weak chemotaxis', meaning small fractional changes in
the tumble rate, \emph{i.e.} $|\Delta(t)| \ll 1$. Physically, weak
chemotaxis corresponds to a shallow chemoattractant gradient.

Brownian rotation causes random re-orientation of the cell swimming
direction, so that in between tumbles the probability density function
$f$ of the swimming direction $\bs{e}$ evolves according to the
Fokker-Planck equation
\begin{equation}
\label{eq:fpe}
   \frac{\partial f}{\partial t}
   + \nabla_{\bs{e}} \cdot [\bm{\omega}(\bs{e}) \times \bs{e} f] =
   \DR \nabla_{\bs{e}}^2 f,
\end{equation}
where $\bm{\omega}(\bs{e})$ is the deterministic angular velocity of
the cell due to the shear flow, $\DR$ is the rotational diffusion
coefficient and $\nabla_{\bs{e}}$ is the gradient operator in direction
space. The form of $\bm{\omega}(\bs{e})$ depends on what assumptions
are made about the cell shape. As discussed in the introduction, the
rotational diffusivity of the cell may be due to intrinsic randomness
in its swimming motion, rather than true Brownian rotation. Throughout
this paper we shall set either $\DR = 0$ or $\DR = 0.062\;{\rm
radians}^2{\rm s}^{-1}$, where the latter value is consistent with what
has been measured in experiments on \emph{E. coli} \citep{Berg.1983}.

When a cell tumbles, its choice of new direction is governed by a
probability distribution which is axisymmetric about the initial
direction. We allow for the tumbles to exhibit directional persistence,
so that the expected scalar product of the swimming directions
$\bs{e}(0^-)$ and $\bs{e}(0^+)$ immediately before and after a tumble
at time $t=0$ is given by
\begin{equation}
\label{eq:alphap}
   E[ \bs{e}(0^-) \cdot \bs{e}(0^+) ] = \ap,
\end{equation}
where $E$ denotes an expectation value and $\ap$ is the `persistence
parameter'. Experimentally, $\ap \approx 0.33$ \citep{Berg.1983}. We
treat tumbles as instantaneous and neglect any rotation of the cell
caused by the shear flow during the tumble.

Our aim is to calculate the chemotactic drift velocity. Let $z_{\rm f}$
be the $z$ location of a cell at the end of a run, relative to its
position at the beginning of the run, and let $t_{\rm f}$ be the
duration of a run. Since we treat tumbles as instantaneous, the
chemotactic drift velocity $\vd$ is
\begin{equation}
\label{eq:vd}
   \vd = E[\zf] / E[\tf].
\end{equation}
While our model assumes an unbounded domain, we note that for a cell
swimming in a bounded domain of length $\gg \vs / \lambda_0$, $\vd$
provides a measure of the transient velocity up the chemoattractant
gradient before the cell encounters the boundaries.

\section{Response function} \label{sec:choosingR}

In writing (\ref{eq:Delta}), we assume that the tumble rate is a linear
functional of the concentration history seen by the cell. The validity
of this assumption has never been directly assessed, and the
chemotactic response function $R$ has never been directly measured.
However, experiments by \citet{Block.1982} and \citet{Segall.1986}
indicate that the bias of a single flagellar motor (the probability of
counter-clockwise rotation) is a linear functional of the concentration
history, thus providing indirect evidence for (\ref{eq:lambda}). These
authors monitored the direction of rotation of a single rotary motor on
an \emph{E. coli} cell while delivering small impulses of
chemoattractant to the fluid around the cell. Repeating the experiment
multiple times and with different cells, the experimenters determined
the motor bias as a function of time (see for example figure 1 of
\citet{Segall.1986}). The impulse response of the motor bias is
double-lobed, with the bias rising above the baseline for the first
$\approx 1\;{\rm s}$ after the delivery of the impulse, falling below
the baseline for the following $\approx 3\;{\rm s}$, and then returning
to baseline. The two lobes of the response have equal area.
Furthermore, it was found that the responses to other time-series of
stimuli (\emph{e.g.} ramp or sinusoidal changes in chemoattractant
concentration) are consistent with the cell behaving as a linear
system, so that the motor bias is well described by the convolution
integral of the stimulus with the impulse response (\emph{c.f.}
equation \ref{eq:lambda}). The primary exception to this linear
behaviour is that for small changes in chemoattractant concentrations,
cells respond to increases in concentration but not to decreases; we
neglect this nonlinearity in our analysis.

\begin{figure}
\centering
  \includegraphics[width=0.7\textwidth]{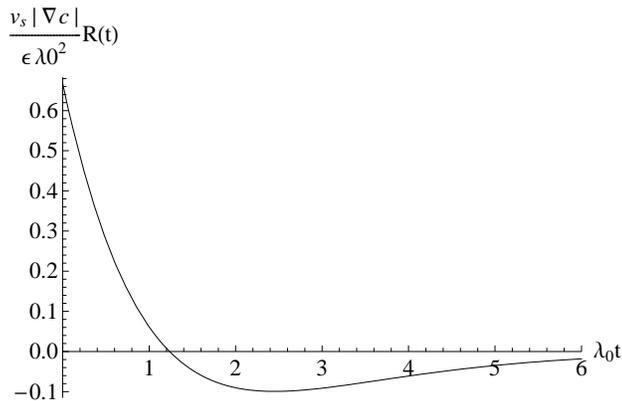}
\caption{The response function used in the results section of this
paper, originally proposed by \citet{Clark.2005}.}
\label{fig:rclarkplot}       
\end{figure}

Following the lead of previous papers
\citep{Schnitzer.1993,deGennes.2004,Clark.2005}, we adopt the view that
in the absence of experimental measurements of $R$, a convenient
assumption is that $R$ has a similar shape to the impulse response of
the individual motor bias reported in \citet{Block.1982} and
\citet{Segall.1986}. This means that (i) $R$ should be composed of a
positive lobe followed by a negative lobe of equal area so that
\begin{equation} \int_0^\infty R(t)\,dt = 0,
\end{equation} and (ii) $R(t)$ should decay to zero for $t$ greater
than about $4\;{\rm s}$. An additional constraint is that $R$ must be
small enough for our linear analysis to be valid. Certainly, $R$ must
be small enough that $\lambda$ never becomes negative, which
necessitates
\begin{equation}
   \epsilon \equiv
   |\Delta(t)|_{\rm max}=
   \vs \nablac \left|
   \int_0^\infty t\, R(t)\,dt \right| < 1.
\end{equation}
The analysis that will be presented in this paper is asymptotically
accurate as $\epsilon \rightarrow 0$.

We shall use the response function
\begin{equation}
\label{eq:clarkR}
   R(t) = \frac{2 \epsilon \lambda_0^2}{3 v_{\rm s} |\nabla c|}
   \,\rm{e}^{-\lambda_0 t}[1-\lambda_0 t/2-(\lambda_0 t/2)^2],
\end{equation}
with $\epsilon \ll 1$ in order for our linear analysis to be valid.
This $R$, plotted in figure \ref{fig:rclarkplot}, is a theoretically
motivated response function proposed by \citet{Clark.2005} that has the
above-mentioned properties (i) and (ii), and it matches the
experimentally measured motor bias reasonably well. It may seem odd
that $|\nabla c|$ appears on the right hand side of (\ref{eq:clarkR}),
since $R$ is a property of the cell rather than its environment.
However the factor of $1/|\nabla c|$ in $R$ simply reflects the fact
that we have used $\epsilon$ to parameterise the strength of the
combined effect of the cell's response $R$ and the chemoattractant
gradient $\nabla c$ on $\Delta(t)$. Note that in the computations we do
not have to specify a value for $|\nabla c|$, since $R$ and $|\nabla
c|$ appear together as a product in the equation for $\Delta(t)$.

\section{Partial calculation of drift velocity $\vd$}
\label{sec:vd}

In this section we present the first part of the analytic drift
velocity calculation, which is common to both the spherical and
non-spherical cell cases. Calculating $\vd$ is non-trivial because of
the interdependence of the tumble rate and the path taken by the cell.
The tumble rate at any time depends in principle on the entire path
history of the cell through (\ref{eq:Delta}), while the path of the
cell depends in turn on the tumble rate. The method of solution is the
same as that used in \citet{Locsei.2007}, so we shall omit some
details. As noted earlier, in order to make the analysis tractable, our
analysis will be restricted to weak chemotaxis, \emph{i.e.}
\mbox{$0\leq|\Delta(t)| \ll 1$}. In this case, the expectation values
of the run displacement $\zf$ and duration $\tf$ satisfy $E[\zf] =
v_{\rm s} \lambda_0 O(\Delta)$ and $E[\tf] = [1+O(\Delta)]/\lambda_0$,
so, from (\ref{eq:vd}),
\begin{equation} \label{eq:vd1}
   \vd = \lambda_0 E[\zf] + v_{\rm s} O(\Delta^2).
\end{equation}
We shall neglect terms that are $O(\Delta^2)$.

Consider a run commencing at time $t = 0$ at location $z = 0$. During
the run, the cell swims in a random walk governed by rotational
Brownian motion until the run terminates with a tumble at time $\tf$.
Note that the expected stopping location of a terminated random walk
with a stopping rate $\lambda_{\rm stop}$ is the same as the expected
first event location on an unterminated walk with an event rate
$\lambda_{\rm event} = \lambda_{\rm stop}$. Thus, in calculating
$E[\zf]$, it is permissible to treat tumbles for $t>0$ as events that
have no effect on the cell's motion, and treat the tumble at $t = \tf$,
$z=\zf$ as a first event (the first tumble in $t>0$). The utility of
this treatment is that we may conceptually break the expectation $E$ in
$E[\zf]$ into two consecutive operations. First, assuming a given path
$z(t):-\infty<t<\infty$ taken by the cell, one calculates the
conditional expectation of $\zf$ for that path. Second, one takes the
expectation over all such paths to obtain $E[\zf]$, with the
understanding that in the $t\leq 0$ section of a path the cell is
subject to reorientations due to both Brownian motion and tumbles,
whereas in the $t>0$ section of a path the cell is subject to
reorientation due to Brownian motion alone. Writing out the two
expectations in symbolic notation,
\begin{equation}
\label{eq:zf0}
   E[ \zf ] = E_{\rm paths}\left[\int_0^{\infty} dt\,
   z_{\rm path}(t) p_{\rm path}(t) \right],
\end{equation}
where the $E_{\rm paths}$ denotes an expectation over paths, $z_{\rm
path}(t)$ denotes the position of the cell at time $t$ on a particular
path, and $p_{\rm path}(t)$ is the probability density function for the
tumble time $\tf$ on a particular path. Since tumbles for $t>0$ are
treated as having no effect on cell motion, paths are independent of
$\tf$ and one is free to take the path expectation inside the integral
over tumble times. Substituting (\ref{eq:zf0}) into (\ref{eq:vd1}) and
dropping the `path' subscript for brevity, one then has
\begin{equation}
\label{eq:vd0}
   \vd = \lambda_0 \int_0^{\infty} dt\, E[z(t) p(t)].
\end{equation}
The probability density function p(t) for the tumble time is given by
\begin{equation}
\label{eq:ptf}
   p(t) = \lambda(t)
   \exp\left[-\int_0^{t} \lambda(t')\,dt'\right],
\end{equation}
where $\lambda(t)$ is the path-dependent tumble rate at time $t$, given
by (\ref{eq:lambda}).

We shall calculate the drift velocity for the case where $R$ is given
by a Dirac delta function,
\begin{equation}
\label{eq:Rdelta}
   R(t)=A\,\delta(t-T),
\end{equation}
and later generalise to an arbitrary response function. Expanding
(\ref{eq:vd0}) in powers of $\Delta$, keeping only the linear term,
then using equation (\ref{eq:Rdelta}) for $R$ and rearranging, one
finds
\begin{equation} \label{eq:vd1new}
   \vd =\lambda_0 \int_0^\infty dt \, {\rm e}^{-\lambda_0 t} E[w(t)]
   + \lambda_0^2 A |\nabla c| \int_0^\infty dt'
   \int_{t'}^\infty dt\, {\rm e}^{-\lambda_0 t} E[w(t) z(t'-T)],
\end{equation}
where $w(t)=d\,z(t)/dt$. Writing $z(t)$ as an integral of $w(t)$,
writing $w(t)$ as $\vs\bs{e}(t).\zhat$, (\ref{eq:vd1new}) becomes
\begin{eqnarray} \label{eq:vd2new}
\nonumber
   \vd =&&\lambda_0 \int_0^\infty dt \, {\rm e}^{-\lambda_0 t}
   \zhat \cdot E[\bs{e}(t)]\\
   &&+ \lambda_0^2 A |\nabla c| \int_0^\infty dt'
   \int_{t'}^\infty dt
   \int_{0}^{t'-T} dt''
   \, {\rm e}^{-\lambda_0 t}
   \zhat \cdot E[\bs{e}(t) \bs{e}(t'')] \cdot \zhat,
\end{eqnarray}

Equation (\ref{eq:vd2new}) contains the unknown terms $E[\bs{e}(t)]$
and $E[\bs{e}(t) \bs{e}(t'')]$. The calculation of these terms under
different sets of assumptions is the subject of the following two
sections.

\section{Analytic calculation of $\vd$ for a spherical organism}

\label{sec:vdspherical}

\subsection{Derivation}

For the case of a spherical organism, one can solve for $\vd$
analytically, taking full account of temporal comparisons, persistence,
and Brownian rotation. Once again, the method of solution is similar to
that in \citet{Locsei.2007}, so just an outline of the derivation is
sketched here.

First consider the case with no shear. In this case, the expected
swimming direction for times $t>0$ is
\begin{equation} \label{eq:e_of_t_noshear}
   E[\bs{e}(t)] = {\rm e}^{-2\DR t} E[\bs{e}(0^+)],
\end{equation}
where $\bs{e}(0^+)$ is the cell's swimming direction at the beginning
of the run \citep{Locsei.2007}. The decaying exponential reflects the
random re-orientation of the cell by Brownian rotation. The time
autocorrelation function for swimming direction is
\begin{equation}
\label{eq:et1et2_noshear}
   E[\bs{e}(t_2) \bs{e}(t_1)] = \left\{
   \begin{array}{rl}
      \frac{1}{3}
      {\rm e}^{2\DR(t_1-t_2)} \bs{1} & \;{\rm if\;} 0<t_1<t_2 \\
      \frac{\ap}{3}
      {\rm e}^{(2\DR+\lambda_0[1-\ap])t_1}
      {\rm e}^{-2\DR t_2}
      \bs{1} & \;{\rm if\;} t_1<0<t_2
   \end{array}
   \right.
\end{equation}
where $\bs{1}$ is the identity matrix. Note that the second case in
(\ref{eq:et1et2_noshear}) has a factor $\ap$ due to the tumble at
$t=0$, an exponential decay term featuring $\DR$ and $\lambda_0$ due to
tumbles and Brownian rotation in the interval $t_1<t<0$, and an
exponential decay term featuring just $\DR$ due to Brownian rotation in
the interval $0<t<t_2$. For Reynolds numbers ${\rm Re} \ll 1$, a
spherical object in a shear flow rotates with an angular velocity
$\bm{\omega}$ equal to half the vorticity \citep{Kim.1992}. When shear
is included, one can show that (\ref{eq:e_of_t_noshear}) and
(\ref{eq:et1et2_noshear}) generalise to
\begin{equation}
   E[\bs{e}(t)] = \rm{e}^{-2\DR t} \bs{R}(t)\cdot E[\bs{e}(0^+)],
\end{equation}
and
\begin{equation}
\label{eq:et1et2}
   E[\bs{e}(t_2) \bs{e}(t_1)] = \left\{
   \begin{array}{rl}
      \frac{1}{3} {\rm e}^{2\DR(t_1-t_2)} \bs{R}(t_2-t_1) & \;{\rm if\;} 0<t_1<t_2 \\
      \frac{\ap}{3}
      {\rm e}^{(2\DR+\lambda_0[1-\ap])t_1}
      {\rm e}^{-2\DR t_2}
      \bs{R}(t_2-t_1) & \;{\rm if\;} t_1<0<t_2
   \end{array}
   \right.
\end{equation}
where $\bs{R}(t)$ is the rotation matrix corresponding to a rotation by
angle $|\bm{\omega}|t$ about the unit vector $\hat{\bm{\omega}}$.

It remains to find an expression for $E[\bs{e}(0^+)]$. Since the
swimming direction after a tumble is chosen from an axisymmetric
distribution around the swimming direction before the tumble,
(\ref{eq:alphap}) gives
\begin{equation}
   E[\bs{e}(0^+)] = \alpha_p E[\bs{e}(0^-)].
\end{equation}
Furthermore, if the cell has been swimming for a time $t_{\rm swim} \gg
1/\lambda_0$, then the expected swimming velocity just before a tumble
does not change from one tumble to the next, so
\begin{eqnarray}
\nonumber
   E[\bs{e}(0^+)] &=& \ap E[\bs{e}(\tf^-)] \\
   &=& \ap \int_0^t dt\, E[\bs{e}(t)p(t)].
\end{eqnarray}
After expanding $p(t)$ in powers of $\Delta$ and performing some
simplification, one finds
\begin{equation}
   E[\bs{e}(0^+)] = \vs \lambda_0 \nablac A \left(\bs{B} - \frac{1}{\ap}\bs{1}\right)^{-1}
   \cdot(\bs{I}_1 - \bs{I}_2)
\end{equation}
where
\begin{equation}
   \bs{I}_1 = \int_0^{\infty} dt \int_0^{(t-T)} dt'
      {\rm e}^{-\lambda_0 t} E[\bs{e}(t)\bs{e}(t')],
\end{equation}
\begin{equation}
   \bs{I}_2 = \lambda_0 \int_0^{\infty} dt' \int_{t'}^{\infty} dt
   \int_0^{t'-T} dt'' \,{\rm e}^{-\lambda_0 t} E[\bs{e}(t)\bs{e}(t'')],
\end{equation}
and
\begin{equation}
   \bs{B} = \lambda_0 \int_0^\infty dt {\rm e}^{-(\lambda_0+2\DR)
   t}\bs{R}(t).
\end{equation}

Now we have expressions for $E[\bs{e}(t)]$ and $E[\bs{e}(t)
\bs{e}(t'')]$, and performing the necessary back-substitutions into
(\ref{eq:vd2new}) yields
\begin{equation}
   \vd = A k_{\rm sphere}(T),
\end{equation}
where
\begin{equation}
\label{eq:k_of_T_sphere}
   k_{\rm sphere}(T)=
   \frac{
      e^{-T (2 \DR+\lambda_0)} \nablac \vs^2 (\ap-1) \lambda_0 b_1
   }{
      3 [(2 \DR+\lambda_0)^2+\Omega ^2]
      [(2 \DR-\ap \lambda_0+\lambda_0)^2+\Omega^2]^2
   },
\end{equation}
with
\begin{equation}
   b_1 = e^{T (2
   \DR+\lambda_0)} \ap \lambda_0 b_2 -e^{T \ap \lambda_0}
   [(2 \DR+\lambda_0)^2+\Omega^2] b_3
\end{equation}
\begin{equation}
   b_2 = (2 \DR+\lambda_0) (2
   \DR-\ap \lambda_0+\lambda_0)^2+[(2 \ap-3) \lambda_0-6
   \DR] \Omega ^2
\end{equation}
\begin{eqnarray}\nonumber
   b_3 =&& [(2 \DR-\ap \lambda_0+\lambda_0)^2-\Omega ^2]
   \cos \Omega T\\
   &&-2 (2 \DR-\ap \lambda_0+\lambda_0) \Omega  \sin \Omega T.
\end{eqnarray}
The drift velocity for a general response function is then
\begin{equation}
\label{eq:vdsimpleintegralsphere}
   \vd = \int_0^\infty dT \, R(T) k_{\rm sphere}(T).
\end{equation}

\subsection{Consistency with known results}
\label{subsec:consistencysphere}

Equations (\ref{eq:k_of_T_sphere}) and
(\ref{eq:vdsimpleintegralsphere}) are consistent with existing
literature in the appropriate limits. \citet{Locsei.2007} considered
chemotaxis in the absence of shear. When $\Omega = 0$,
(\ref{eq:k_of_T_sphere}) simplifies to
\begin{equation}
\label{eq:k_of_T_noshear}
   k(T) =
   \frac{
      v_{\rm s}^2 |\nabla c| \lambda_0
      {\rm e}^{-(\lambda_0+2\DR)T}(1-\ap)
      [(\lambda_0+2\DR){\rm e}^{\lambda_0 \ap T}
      -\lambda_0 \ap {\rm e}^{(\lambda_0+2\DR)T}]
   }{
      3(\lambda_0+2\DR)[\lambda_0(1-\ap)+2\DR]^2
   },
\end{equation}
consistent with \citep{Locsei.2007}. \citet{Bearon.2000} calculated the
chemotactic drift velocity for a spherical cell swimming in a uniform
shear, under the assumption that the cell detects the chemoattractant
gradient directly (rather than through temporal comparisons) and
modifies its tumble rate according to
\begin{equation}
\label{eq:lambdabearon}
   \lambda=\lambda_0(1 - \epsilon \bs{e} \cdot \zhat),
\end{equation}
with $|\epsilon| \ll 1$. We note that within our framework
(\ref{eq:lambdabearon}) is equivalent to setting
\begin{equation}
\label{eq:Rbearon}
   R(t) = \frac{\epsilon}{\vs \nablac}
   \frac{\delta(t-T)-\delta(t-T-\Delta t)}{\Delta t},
\end{equation}
and taking the simultaneous limits $\Delta t \rightarrow 0$ and $T
\rightarrow 0$. Substituting (\ref{eq:Rbearon}) into
(\ref{eq:vdsimpleintegralsphere}) and taking the aforementioned limits,
one finds the drift velocity for a gradient-detecting spherical
organism:
\begin{equation} \label{eq:vdgradientdetection}
   \vd = \frac{
      \epsilon \vs \lambda_0 (\ap - 1)[\lambda_0(\ap-1)-2\DR]
   }{
      3[(\lambda_0 (\ap -1)-2\DR)^2+\Omega^2].
   }
\end{equation}
Setting $\ap = 0$ and $\DR = 0$, one finds
\begin{equation} \label{eq:vdBearon}
   \vd = \frac{\epsilon \vs \lambda_0^2}{3(\lambda_0^2+\Omega^2)},
\end{equation}
consistent with equation 30 of \citet{Bearon.2000}.

\subsection{Results: $\vd$ for a spherical organism}
\label{subsec:results_sphere}

Using response function (\ref{eq:clarkR}), the drift velocity for a
spherical organism in a shear flow is
\begin{equation} \label{eq:vdsphere}
   \frac{\vd}{\epsilon \vs} = \frac{(\ap-1) \lambda_0^3 \{b_5 [2 \DR-(\ap-2)
   \lambda_0] \Omega ^2-b_4+b_6\}}{b_7}
\end{equation}
where
\begin{equation}
   b_4 = [4 \DR+(5-2 \ap) \lambda_0] [2 \DR-(\ap-2) \lambda_0]^3 (2
   \DR-\ap \lambda_0+\lambda_0),
\end{equation}
\begin{equation}
   b_5 = 16 \DR^2+4 (11-4 \ap) \lambda_0 \DR+\left(4 \ap^2-22
   \ap+25\right) \lambda_0^2,
\end{equation}
\begin{equation}
   b_6 = 3 [4 \DR+(3-2 \ap) \lambda_0] \Omega
   ^4,
\end{equation}
and
\begin{equation}
   b_7 = 9 \{[(\ap-2) \lambda_0-2 \DR]^2+\Omega ^2\}^3 [(2 \DR-
   \ap \lambda_0+\lambda_0)^2+\Omega ^2].
\end{equation}

\begin{figure}
\centering
  \includegraphics[width=0.7\textwidth]{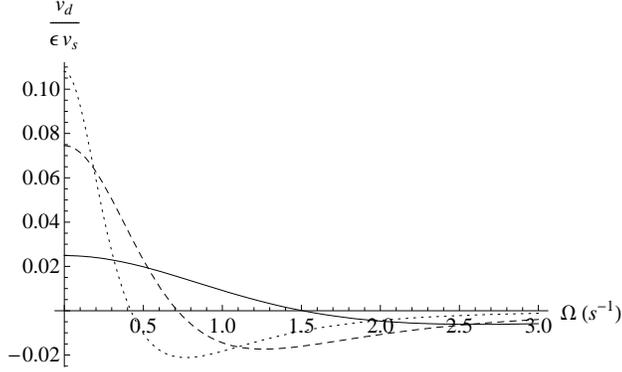}
\caption{Drift velocity versus shear strength, for persistence
parameter $\ap=-1$ (solid line), $\ap=0.33$ (dashed), and $\ap=0.78$
(dotted), with $\lambda_0 = 1\;{\rm s}^{-1}$ and $\DR = 0.062 \;{\rm
radians}^2{\rm s}^{-1}$. Note $\ap = 0.33$ is the measured value for
\emph{E. coli} \citep{Berg.1983}.}
\label{fig:vd_sphere_vs_Omega}       
\end{figure}

Figure \ref{fig:vd_sphere_vs_Omega} shows the drift velocity $\vd$ as a
function of the shear strength $\Omega$ for a variety of persistence
parameters $\ap$. As one might expect, the shear flow reduces the drift
velocity. More surprisingly, the drift velocity is negative for
sufficiently strong shear, so that the cell swims down the
chemoattractant gradient rather than up it. Also, the larger the
persistence parameter, the more susceptible the cell is to shear. For
instance, with a persistence parameter of $\ap=-1$, $\vd$ becomes
negative for $\Omega$ beyond $\approx 1.5\;{\rm s}^{-1}$, whereas for
$\ap = 0.78$, the drift velocity becomes negative for $\Omega$ beyond
$\approx 0.4\;{\rm s}^{-1}$. \citet{Locsei.2007} calculated that in the
absence of shear, $\ap \approx 0.78$ is the value of persistence that
maximises the drift velocity. In that paper, it was speculated that the
measured persistence of $\ap \approx 0.33<0.78$ in real \emph{E. coli}
\citep{Berg.1983} might reflect a compromise between transient and
steady state performance. The new results here suggest another reason
why $\ap < 0.78$ for real \emph{E. coli}: there is a compromise between
performance in a stationary fluid and performance in a sheared fluid.
Intuitively, the disadvantage of a high persistence parameter when
performing chemotaxis in a sheared fluid arises because when a cell is
rotated by the shear to face down the chemoattractant gradient, a
single tumble is insufficient to reorient it to face up the gradient.

\section{Semi-analytic calculation of $\vd$ for an elongated organism}

\label{sec:vdspheroid}

Whereas a spherical particle in a shear flow rotates with constant
angular velocity, a non-spherical particle rotates with an angular
velocity that depends on its instantaneous orientation. This
complicates the calculation of drift velocity, but nonetheless one can
still find an analytic integral expression for the drift velocity of a
non-spherical cell in the absence of Brownian rotation and persistence,
using the additional simplification of treating the cell as a prolate
spheroid. In principle, one could include Brownian rotation and
persistence in the calculation, but this would involve solving the
full, time-dependent Fokker Plank equation, and we shall not attempt
this here.

\subsection{Modelling an \emph{E. coli} cell as a prolate spheroid}
\label{subsec:spheroid}

In this section we show that an \emph{E. coli} cell in a shear flow
rotates in approximately the same manner as a prolate spheroid
(ellipsoidal rod) of slenderness ratio $\eta \approx 9$, where $\eta$
is defined as the ratio of major to minor axis.

We begin by considering the motion of a rigid body in a linear velocity
field. A general, undisturbed linear velocity field $\bs{u}(\bs{x})$
has the form
\begin{equation}
\label{eq:linflow}
   \bs{u}(\bs{x}) = \bs{U} + \bm{\Omega} \times \bs{x} + \bs{E}\cdot
   \bs{x},
\end{equation}
where $\bs{U}$ is a uniform velocity, $\bm{\Omega}$ is an angular
velocity equal to half the vorticity, and $\bs{E}$ is a second rank
symmetric strain rate tensor. The shear flow considered in this paper
is a special case of a linear flow, with
\begin{eqnarray} \label{eq:Omega}
\nonumber
   \bs{U}&=&\bs{0}, \\
\nonumber
   \bm{\Omega} &=& (0,\Omega,0), \\
   \bs{E}&=& \left(
   \begin{array}{ccc}
      0 & 0 & \Omega \\
      0 & 0 & 0 \\
      \Omega & 0 & 0 \\
   \end{array}
   \right),
\end{eqnarray}
where $\Omega = \gamma / 2$.

A swimming \emph{E. coli} cell has a Reynolds number of ${\rm Re}
\approx 10^{-5}$ \citep{Berg.1983} so the hydrodynamics of its motion
are well described by Stokes flow. By the linearity properties of
Stokes flow, the instantaneous angular velocity $\bm{\omega}$ of any
rigid body in a linear flow takes the form
\begin{equation} \label{eq:omegai}
   \omega_i = \Omega_i + B_{ijk} E_{jk}
\end{equation}
(using the Einstein summation convention), where $B_{ijk}$ is a
dimensionless third rank constant tensor which is symmetric in its
second and third indices, and the $i$, $j$, and $k$ components refer to
axes fixed in the body. For a sphere, $\bs{B} = \bs{0}$, reflecting the
fact that a sphere does not rotate in a pure straining flow.
\citet{Bretherton.1962} showed that for a general axisymmetric body
$B_{ijk}$ has only a single free parameter. Taking the $z$ axis to be
aligned with the body's symmetry axis, an axisymmetric body has
\begin{eqnarray} \label{eq:Baxisym}
\nonumber
   -B_{123} = -B_{132}
   = B_{213} = B_{231}=\alpha,\\
   B_{ijk}=0 \textrm{ for all other } i,j,\,k.
\end{eqnarray}

\citet{Bretherton.1962} further showed that, provided $|\alpha|<1$
(which is true for most physically realistic bodies), a body with
$B_{ijk}$ of the form (\ref{eq:Baxisym}) rotates in a shear flow in
precisely the same way as a spheroid of slenderness ratio
\begin{equation} \label{eq:equiveta}
   \eta = \sqrt{(1+\alpha)/(1-\alpha)}.
\end{equation}
This result is useful because the motion of a prolate spheroid in a
shear flow was analysed by \citet{Jeffery.1922}, who showed that the
motion consists of a closed orbit of simple analytical form, now
commonly referred to as a `Jeffery orbit'.

We shall consider a model \emph{E. coli} cell composed of a spherical
body rigidly attached to a helical flagellar bundle, modelled as a
solid body of uniform circular cross-section, and show that despite the
model cell being non-axisymmetric its $B_{ijk}$ tensor is almost of the
form (\ref{eq:Baxisym}). Our model cell consists of a sphere of radius
$a$ centered at the origin with respect to axes fixed in the cell and a
left-handed helix of radius $r$, pitch (helical wavelength) $h$ and
length $L$, such that the position $\bs{x}$ at a point on the helix is
given parametrically by
\begin{equation}
   \bs{x}(\chi)=r(\bs{n}_1 \cos\chi - \bs{n}_2 \sin\chi)
      + \bs{n}_3\left(a+\frac{h \chi}{2\pi}\right),
\end{equation}
where $\chi \in [0,2\pi L/h]$ and $\bs{n}_1,\bs{n}_2,\bs{n}_3$ are a
right-handed set of unit vectors fixed in the cell such that $\bs{n}_3$
points along the helix axis away from the sphere.

It is sufficient to treat the helix and sphere as being fixed in
position relative to one another, so that the cell is rigid rather than
swimming. By the linearity property of Stokes flow, the angular
velocity of a swimming cell in a shear flow is equal to the angular
velocity of a swimming cell in a stationary fluid, plus the angular
velocity of a rigid cell in a shear flow. Since
$\Omega_1=\Omega_2\approx0$ for a swimming cell in a stationary fluid
(because the cell swims in an approximately straight line), $\Omega_1$
and $\Omega_2$ are the same for a swimming cell in a shear flow as for
a rigid cell in a shear flow. $\Omega_3$ is not relevant to
calculations of drift velocity, since rotation about $\bs{n}_3$ does
not alter the swimming direction.

For a given background flow, the translational velocity $\bs{V}$ and
angular velocity $\bm{\omega}$ of the body are found by setting the net
force and net torque on the body to zero. We neglect the interaction of
flows around the helix and the sphere. For a given motion of the cell
through a given background velocity field, we assume that the total
viscous force on the cell is simply the force on the sphere in the
absence of the helix plus the force on the helix in the absence of the
sphere, and we similarly assume simple addition of the viscous torques.
This is not an unreasonable approximation since the body diameter
(1--2$\mu{\rm m}$) is a small fraction of flagellar bundle length
(5--10$\mu{\rm m}$).  To calculate the force and torque on the sphere,
we use the well known exact expressions \cite[see for
instance][]{Batchelor.1967}. To calculate the force and torque on the
helix we use resistive force theory \citep{Gray.1955}, which assumes
that the components of the viscous force on an element of the helix are
proportional to the same components of the fluid's velocity relative to
that element, but with different coefficients of proportionality for
normal and tangential components. Various pairs of coefficients have
been proposed for a simple flagellum of uniform circular cross-section,
and we use the more accurate coefficients due to \citet{Lighthill.1976}
rather than the original coefficients proposed by \citet{Gray.1955}.
These coefficients have as parameters the helix pitch $h$ and the
cross-sectional radius $b$ of the helix bundle. Resistive force theory
is a crude approximation of true flagellar hydrodynamics, since it
ignores interactions between neighbouring elements of the flagellum,
but it has nonetheless been shown to give reasonable results when
applied to swimming spermatozoa, for example \citep{Lighthill.1976}.

Given the above method for calculating the motion of the body in a
given flow, one can determine the $B_{ijk}$ tensor from
(\ref{eq:omegai}) by setting $\bm{\Omega} = \bs{0}$ and considering the
angular velocity $\bm{\omega}$ that results from each of the following
elementary strain matrices, in which all $E_{ij}$ except those
mentioned are zero: $E_{11} = 1$; $E_{22}  =1$; $E_{33} = 1$; $E_{12} =
E_{21} = 1$; $E_{23} = E_{32} = 1$; $E_{31} = E_{13} = 1$.

In principle one can determine $B_{ijk}$ in symbolic form with $a$,
$b$, $r$, $L$, and $h$ as parameters, but in practice the algebra
required to perform this task is unwieldy and so one must choose
numerical values for the parameters before calculating $B_{ijk}$. A
typical \emph{E. coli} cell body is rod shaped with a diameter of
$\approx 1\mu {\rm m}$ and a length of $\approx 2\mu {\rm m}$, and
possesses about $6$ flagella \citep{Berg.1983}. The length of each
flagellum is typically in the range 5--10$\mu{\rm m}$, and the diameter
of each flagellar filament is $\approx 20{\rm nm}$ \citep[pp.
70--83]{Neidhardt.1987}. The pitch of the helix is $\approx 2.5 \mu{\rm
m}$ and the helix radius is $\approx 0.25 \mu {\rm m}$
\citep{Turner.2000}. On the basis of these measurements, we take the
dimensions of our model cell to be $a = 0.7 \mu {\rm m}$, $r = 0.25 \mu
{\rm m}$, $L = 8 \mu {\rm m}$, $h = 2.5 \mu {\rm m}$, $b = \sqrt{6}
\times 10 {\rm nm}$. With these dimensions, calculation of the
$B_{ijk}$ reveals that, for the model cell, they come close to
possessing the properties expected for an axisymmetric body [eq.
(\ref{eq:Baxisym})], in that
\begin{eqnarray}
\nonumber
   -B_{123} = -B_{132}
   \approx B_{213} = B_{231} \approx 0.974,\\
\nonumber
   B_{333} \approx 0.501, \\
   |B_{ijk}|<0.05 \textrm{ for all other } i,j,\,k.
\end{eqnarray}
The most notable breaking of symmetry is that $B_{333}$ is not small
(whereas it is zero for an axisymmetric body) but this only affects
$\Omega_3$ and thus has no bearing on the swimming direction. All other
elements of $B_{ijk}$ that are zero for an axisymmetric body have
magnitude $<0.05$ for the model cell (and all but one, $B_{133} =
-0.0445$, have magnitude $\leq 0.01$), so discrepancies between the
angular velocities of the model cell and an `equivalent spheroid' are
of order $5\%$. Using (\ref{eq:equiveta}), the slenderness ratio of the
`equivalent spheroid' that rotates in a similar manner to the model
cell is $\eta \approx 9$. For other plausible values of $a$ and $L$,
$\eta$ ranges from $\eta \approx 4$ (for $a=1\mu {\rm m}$, $L=5\mu {\rm
m}$) to $\eta \approx 15$ (for $a=0.5\mu {\rm m}$, $L = 10\mu {\rm
m}$).

\subsection{Integral expression for drift velocity of prolate spheroid}
\label{subsec:autocorrlnspheroid}

Neglecting Brownian rotation and persistence, one can find an analytic
integral expression for the drift velocity. The direction faced at the
beginning of each run is taken from the uniform distribution on the
unit sphere, so by symmetry $E[\bs{e}(t)] = \bs{0}$ and the first term
in (\ref{eq:vd2new}) vanishes. Also, since any correlation in swimming
direction is destroyed by the tumble, $E[\bs{e}(t)\bs{e}(t'')]=\bs{0}$
for $t''<0<t$, which allows some simplification of the second term in
(\ref{eq:vd2new}). One finds
\begin{equation}
   \vd =
   \frac{\lambda_0^2 A \nablac}{\vs^2}
   \int_T^\infty dt\, {\rm e}^{-\lambda_0 t}
   \int_0^{(t-T)} dt'\,
   (t-t'-T)E[w(t)w(t')].
\end{equation}
It remains to find an expression for the velocity autocorrelation
function $E[w(t)w(t')]$.

Since it was established in section \ref{subsec:spheroid} that an
\emph{E. coli} cell rotates in a shear flow with very nearly the same
angular velocity as a prolate spheroid, we henceforth treat the cell as
a prolate spheroid that swims along its long axis. Define direction
angles $\theta$ and $\phi$ such that
\begin{eqnarray}
\label{eq:edotez}
   \bs{e} \cdot \zhat &=& \sin\theta \,
   \cos\phi \\
   \bs{e} \cdot \hat{\bs{x}} &=& \sin\theta \, \sin\phi \\
   \bs{e} \cdot \hat{\bs{y}} &=& \cos\theta.
\end{eqnarray}
In a shear flow, neglecting Brownian rotation, the cell follows Jeffery
orbits, and the angles $\theta$ and $\phi$ evolve according to
\citep{Jeffery.1922}
\begin{eqnarray}
\label{eq:tanphi}
   \tan\phi &=& \eta \tan(\sigma t-\beta_0), \\
\label{eq:tansqtheta}
   \tan^2 \theta &=& \frac{\eta^2}{\kappa_0(\eta^2\cos^2\phi+\sin^2\phi)},
\end{eqnarray}
where $\kappa_0$ and $\beta_0$ are constants of integration depending
on the initial orientation, and
\begin{equation} \label{eq:sigma}
   \sigma=\frac{2\Omega \, \eta}{\eta^2+1}.
\end{equation}
In terms of the initial orientation angles $\theta_0$ and $\phi_0$ at
time $t=0$, the constants of integration are
\begin{eqnarray}
\label{eq:beta0}
   \beta_0(\phi_0) &=& - \arctan(\eta \cos\phi_0, \sin\phi_0), \\
\label{eq:k0}
   \kappa_0(\theta_0,\phi_0) &=&
   \frac{\eta^2}{\tan^2\theta_0(\eta^2\cos^2\phi_0+\sin^2\phi_0)},
\end{eqnarray}
where the $\arctan$ function is treated as taking two arguments so as
to return an answer in the correct quadrant. For instance, if $\psi =
\arctan(x_1,x_2)$, then $\cos\psi = x_1(x_1^2+x_1^2)^{-1/2}$ and
$\sin\psi = x_2(x_1^2+x_1^2)^{-1/2}$.

Let $w(t;\eo)$ be the cell's velocity projected along the $z$ axis at a
time $t\geq0$, given that the cell's swimming direction at the
beginning of the run was $\eo$. Manipulation of (\ref{eq:edotez}),
(\ref{eq:tanphi}) and (\ref{eq:tansqtheta}) gives
\begin{eqnarray}
\label{eq:ez} \nonumber
   w(t;\eo)&=& \vs \sin\theta\cos\phi\\
   &=&\frac{\vs \cos(\sigma t-\beta_0)}
         {\left[{\kappa_0+\cos^2(\sigma t-\beta_0)+\eta^2\sin^2(\sigma
         t-\beta_0)}\right]^{1/2}}.
\end{eqnarray}
$E[w(t_1)w(t_2)]$ may be written as an explicit average over all
possible initial run directions:
\begin{equation}
\label{eq:correlation}
   E[w(t_1)w(t_2)] = \frac{1}{4 \pi}
   \int_0^{\pi}d\theta_0 \sin\theta_0
   \int_0^{2\pi}d\phi_0 w(t_1;\eo) w(t_2;\eo),
\end{equation}
where the integrand is treated as a function of $\theta_0$ and
$\phi_0$. Using (\ref{eq:beta0}), we can transform this last equation
into to
\begin{equation}
\label{eq:correlation2}
   E[w(t_1)w(t_2)] = \frac{\eta}{4 \pi}
   \int_0^{\pi}d\theta_0 \sin\theta_0
   \int_0^{2\pi}d\beta_0
   \frac{
      w(t_1;\eo)w(t_2;\eo)
   }{
      \cos^2\beta_0+\eta^2\sin^2\beta_0
   },
\end{equation}
where the integrand is now treated as a function of $\theta_0$ and
$\beta_0$, and there is an accompanying expression for $\kappa_0$ in
terms of $\theta_0$ and $\beta_0$ instead of $\theta_0$ and $\phi_0$:
\begin{equation}
\label{eq:k02}
   \kappa_0(\theta_0,\beta_0)=\frac{\cos^2\beta_0+\eta^2\sin^2\beta_0}{\tan^2\theta_0}.
\end{equation}
Equations (\ref{eq:ez}), (\ref{eq:correlation2}), and (\ref{eq:k02})
now give a complete expression for $E[w(t_1)w(t_2)]$. The right hand
side of (\ref{eq:correlation2}) cannot be integrated analytically for
general $\eta$, but can be evaluated by numerical quadrature. In the
limit of weak shear, one can expand the integrand in powers of
$\Omega/\lambda_0$, taking $t_1$ and $t_2$ to be of order
$1/\lambda_0$, and integrate to find
\begin{eqnarray} \label{eq:smallshearC}
\nonumber
   E[w(t_1)w(t_2)] = \frac{\vs^2}{3}
   -\frac{2\vs^2\Omega^2}{105 (\eta ^2+1)^2}
   \{-2 (3 \eta ^4+8 \eta ^2+24) t_1 t_2 \\
   +(12 \eta ^4+11 \eta ^2+12) (t_1^2+t_2^2)
   \}+O(\Omega^4/\lambda_0^4).
\end{eqnarray}
Note that (\ref{eq:smallshearC}) contains only even powers of $\Omega$
since by symmetry the drift velocity does not depend on the sign of
$\Omega$.

For a general response function, the drift velocity is given by
\begin{equation}
\label{eq:vdsimpleintegralspheroid}
   \vd = \int_0^\infty dT \, R(T) k_{\rm spheroid}(T),
\end{equation}
where
\begin{equation} \label{eq:k_of_T_spheroid}
   k_{\rm spheroid}(T) = \frac{\lambda_0^2 \nablac}{\vs^2}
   \int_T^\infty dt\, {\rm e}^{-\lambda_0 t}
   \int_0^{(t-T)} dt'\,
   (t-t'-T)E[w(t)w(t')],
\end{equation}
with $E[w(t)w(t')]$ given by (\ref{eq:correlation2}).

\subsection{Consistency with known results}

\label{subsec:consistencyspheroid}

\citet{deGennes.2004} calculated the chemotactic drift velocity in the
absence of a shear flow, and without Brownian rotation or persistence.
Setting $\Omega = 0$, the velocity autocorrelation function is simply
$E[w(t_1),w(t_2)]=\vs^2/3$, and (\ref{eq:k_of_T_spheroid}) and
(\ref{eq:vdsimpleintegralspheroid}) reduce to
\begin{equation}
   \vd = \frac{\vs^2 \nablac}{3 \lambda_0}
   \int_0^\infty R(t) {\rm e}^{-\lambda_0 t} dt,
\end{equation}
consistent with equation 16 of \citet{deGennes.2004}. As mentioned in
section \ref{subsec:consistencysphere}, \citet{Bearon.2000} calculated
the chemotactic drift velocity for a spherical cell swimming in a
uniform shear, under the assumption that the cell detects the
chemoattractant gradient and modifies its tumble rate according to
(\ref{eq:lambdabearon}). We can treat this case as follows. For a
sphere, $\eta = 1$, and the velocity autocorrelation function
simplifies to
\begin{equation}
\label{eq:Csphere}
   E[w(t_1),w(t_2)]=\frac{\vs^2}{3}\cos[\Omega(t_1-t_2)],
\end{equation}
and substitution of (\ref{eq:Csphere}) into (\ref{eq:k_of_T_spheroid})
yields
\begin{equation} \label{eq:kofTsphere}
   k(T)=-
   \frac{
      {\rm e}^{-\lambda_0 T}[(\Omega^2-\lambda_0^2)\cos\Omega T+2\Omega\sin\Omega T]
   }{
      3(\lambda_0^2+\Omega^2)^2
   }.
\end{equation}
As in section \ref{subsec:consistencysphere}, gradient detection is
modelled by using the response function (\ref{eq:Rbearon}) and taking
the appropriate limits, and one finds that once again the drift
velocity reduces to (\ref{eq:vdBearon}), consistent with
\citep{Bearon.2000}.

\subsection{Results: $\vd$ for an elongated organism}
\label{subsec:results_spheroid}

For weak shear ($\Omega \ll \lambda_0$), we can substitute
(\ref{eq:smallshearC}) and (\ref{eq:clarkR}) into
(\ref{eq:k_of_T_spheroid}) and (\ref{eq:vdsimpleintegralspheroid}) to
find
\begin{equation} \label{eq:vdsmallshear}
   \frac{\vd}{\epsilon \vs}=
   \epsilon \left[
   \frac{5}{72}- \frac{\Omega^2}{\lambda_0^2} g(\eta)
   +O\left(\frac{\Omega^4}{\lambda_0^4}\right)
   \right]
\end{equation}
where
\begin{equation}
   g(\eta) = \frac{243\eta^4+151\eta^2-44}{420(1+\eta^2)^2}.
\end{equation}
Note that $g(\eta)$ is bounded and monotonically increasing for
$\eta>0$, with $g(\infty) = 81/140$. In fact, $g(4) \approx 0.92
\,g(\infty)$, indicating that for $\Omega\ll1$ and $\eta>4$, $\vd$ is
approximately independent of $\eta$.

\begin{figure}
\centering
  \includegraphics[width=0.7\textwidth]{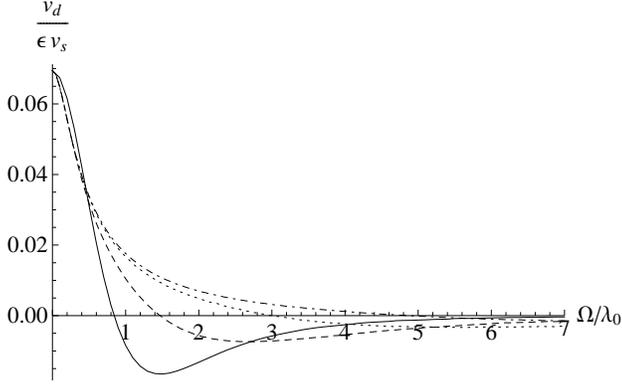}
\caption{Drift velocity $\vd$ as a function of the shear strength
parameter $\Omega$ for equivalent slenderness ratios $\eta = 1$ (solid
line), $\eta = 4$ (dashed line), $\eta = 9$ (dotted line), and $\eta =
15$ (dot-dashed line).}
\label{fig:cubacombiplot}       
\end{figure}

For general $\Omega$ and $\eta$, no analytic simplifications are
possible and we evaluate the integral
(\ref{eq:vdsimpleintegralspheroid}) numerically using the `Cuba'
integration library developed by \citet{Hahn.2005}.
Figure~\ref{fig:cubacombiplot} shows the drift velocity $\vd$ as a
function of the shear strength $\Omega$ for a range of values of the
slenderness ratio $\eta$. Note that for $\Omega < 0.5$ and $\eta \geq
4$, $\vd$ is approximately independent of $\eta$, consistent with the
analysis in the previous paragraph. Also, for any given $\eta$, and for
sufficiently large $\Omega$, the cell exhibits `negative chemotaxis'
\emph{i.e. } $\vd$ is negative. For instance, if $\eta = 9$ then $\vd <
0$ for $\Omega > 3$. The minimum value of $\Omega$ for which $\vd<0$
depends on $\eta$; for larger $\eta$, $\vd$ remains positive up to a
larger value of $\Omega$.

This last finding can be explained as follows. A key feature of the
Jeffery orbit of a prolate spheroid with large slenderness ratio is
that that there are lengthy `pauses' with the spheroid's symmetry axis
almost parallel to the streamlines, but the orientation reverses
periodically. From (\ref{eq:tanphi}) and (\ref{eq:sigma}) we see that
the period $T_{\rm Jef.}$ of a Jeffery orbit is
\begin{equation}
   T_{\rm Jef.} = \pi (\eta^2+1)/(\Omega \eta),
\end{equation}
which is a monotonically increasing function of $\eta$ for $\eta > 1$.
Let $\Omega_{\rm crit}$ be the lowest value of $\Omega$ for which $\vd
= 0$. In order for $\vd$ to be less than or equal to zero, the sign of
$w$ must change during a typical run, and thus $T_{\rm Jef.}$ must be
comparable to or smaller than the mean run duration $1/\lambda_0$.
Thus, we expect that the dependence of $\Omega_{\rm crit}$ on $\eta$
scales roughly as
\begin{equation} \label{eq:omegacrit}
   \Omega_{\rm crit}/\lambda_0 \sim q(\eta) \equiv (\eta^2+1)/\eta.
\end{equation}
In fact, we find that $(\Omega_{\rm crit}/\lambda_0)/q(\eta) = 0.42,
0.35, 0.33, 0.32$ for $\eta = 1, 4, 8, 16$ respectively, indicating
that the scaling (\ref{eq:omegacrit}) is approximately correct.

\section{Calculation of $\vd$ by Monte Carlo simulation}
\label{sec:vdsimulation}

The purpose of using Monte Carlo simulation here is twofold. First, it
gives a means of checking the analytic results derived in earlier
sections. Second, it allows us to calculate $\vd$ in the regime which
was not analytically tractable, namely the regime including all four
effects of (i) temporal comparisons, (ii) persistence, (iii) Brownian
rotation, and (iv) elongated cell shape. One can then check that the
conclusions drawn from the analytic calculations are still valid when
all four effects are taken into account.

\subsection{Monte Carlo algorithm and test cases}

\begin{figure}
\centering
  \includegraphics[width=0.7\textwidth]{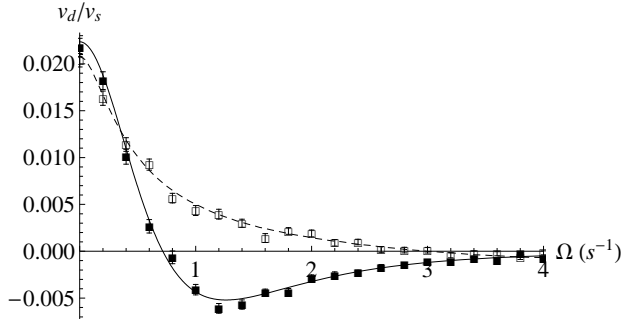}
\caption{Examples of agreement between analytic theory and Monte Carlo
simulation results. Solid line shows analytic result for a spherical
organism with $\ap = 0.33$, $\DR = 0.062 \;{\rm radians}^2{\rm
s}^{-1}$, $\epsilon = 0.3$, and $\lambda_0 = 1\;{\rm s}^{-1}$; solid
squares are Monte Carlo results for the same set of parameters. Dashed
line shows analytic result for an elongated organism with $\eta = 9$,
$\ap = 0$, $\DR = 0 \;{\rm radians}^2{\rm s}^{-1}$, $\epsilon = 0.3$,
and $\lambda_0 = 1\;{\rm s}^{-1}$; empty squares show Monte Carlo
results for the same set of parameters. Error bars show the standard
error of the mean for Monte Carlo results.}
\label{fig:sims_and_theory}       
\end{figure}

The algorithm for calculating $\vd$ by Monte Carlo simulation is
straightforward. For each set of parameter values, 100 simulations are
run, each of duration $10^4/\lambda_0$. Time is discretised into steps
of size $\Delta t = 0.01/\lambda_0$. At each time step, the tumble
probability (\ref{eq:lambda}) is evaluated by numerical quadrature, and
a random number generator is used to decide whether the cell tumbles.
If the cell tumbles, then the new direction is chosen from an
axisymmetric distribution about the old direction, such that the new
direction makes an angle of $\arccos(\ap)$ with the old direction. If
the cell doesn't tumble, then Brownian rotation is simulated by giving
the cell a new direction chosen from an axisymmetric distribution about
the old direction, such that the new direction makes an angle of
$\arccos(1-2\DR\Delta t)\approx 2 \sqrt{\DR\Delta t}$ with the old
direction, and in addition the cell's position and direction of motion
is incremented according to the Jeffery orbit equations
\citep{Jeffery.1922}. A computer code was written to implement the
above algorithm, and its output was compared to the analytic results
derived in earlier sections. As show in figure
\ref{fig:sims_and_theory}, the analytic and simulation results agree
well.

\subsection{The effect of persistence on drift velocity}

\begin{figure}
\centering
  \includegraphics[width=0.7\textwidth]{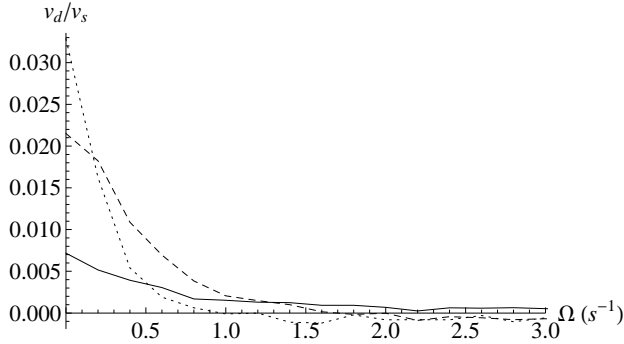}
\caption{Monte Carlo results for drift velocity $\vd$ as a function of
shear strength $\Omega$, for persistence parameters of $\ap = -1$
(solid line), $\ap = 0.33$ (dashed), and $\ap = 0.78$ (dotted). In each
case, $\eta = 9$, $\DR = 0.062\;{\rm radians}^2{\rm s}^{-1}$,
$\lambda_0 = 1\;{\rm s}^{-1}$ and $\epsilon = 0.3$. Standard errors
(not shown) are of the same magnitude as those in figure
\ref{fig:sims_and_theory}.}
\label{fig:simplots_effect_of_persistence}       
\end{figure}

In section \ref{subsec:results_sphere}, it was shown that for a
spherical organism, a higher persistence $\ap$ results in a greater
drift velocity $\vd$ in the absence of shear, but a lower drift
velocity in higher values of shear (figure
\ref{fig:vd_sphere_vs_Omega}). Using the Monte Carlo code, the same
calculation was repeated, but for an elongated organism ($\eta = 9$).
As shown in figure \ref{fig:simplots_effect_of_persistence}, the
results remain qualitatively similar. For instance, $\ap = 0.78$ gives
a higher $\vd$ than $\ap = 0.33$ for shear rates up to $\Omega \approx
0.2\;{\rm s}^{-1}$, but for shear rates beyond that the situation is
reversed. The main difference between figures
\ref{fig:vd_sphere_vs_Omega} and
\ref{fig:simplots_effect_of_persistence} is that when elongated cell
shape is taken into account, $\vd$ never becomes as strongly negative
as it does in the case of a spherical organism.

\subsection{The effect of organism shape on drift velocity}

\begin{figure}
\centering
  \includegraphics[width=0.7\textwidth]{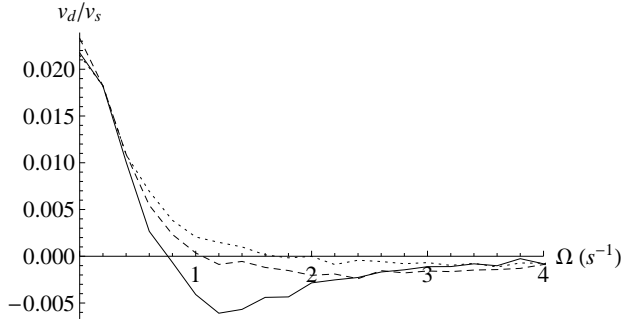}
\caption{Monte Carlo results for drift velocity $\vd$ as a function of
shear strength $\Omega$, for slenderness ratios of $\eta = 1$ (solid
line), $\eta = 4$ (dashed), and $\eta = 9$ (dotted). In each case, $\ap
= 0.33$, $\DR = 0.062\;{\rm radians}^2{\rm s}^{-1}$, $\lambda_0 =
1\;{\rm s}^{-1}$ and $\epsilon = 0.3$. Standard errors (not shown) are
of the same magnitude as those in figure \ref{fig:sims_and_theory}.
Results for $\eta = 15$ are statistically indistinguishable from
results for $\eta = 9$ (data not shown).}
\label{fig:simplots_effect_of_elongation}       
\end{figure}

In section \ref{subsec:results_spheroid}, it was shown that a more
elongated organism (larger $\eta$) is able to perform chemotaxis in
stronger shear flows than a less elongated organism (figure
\ref{fig:cubacombiplot}), in the case where persistence and Brownian
rotation are neglected. Using the Monte Carlo code, the same
calculation was repeated, but including persistence ($\ap = 0.33$) and
Brownian rotation ($\DR = 0.062\;{\rm radians}^2{\rm s}^{-1}$). As
shown in figure \ref{fig:simplots_effect_of_elongation}, the results
remain qualitatively unchanged. For instance, for $\eta = 1$, $\vd$
remains positive for shear rates up to $\Omega \approx 0.8\; {\rm
s}^{-1}$, whereas for $\eta = 9$, $\vd$ remains positive for shear
rates up to $\Omega \approx 1.5 {\rm s}^{-1}$. However, the effect of
$\eta$ is less pronounced than in the case where persistence and
Brownian rotation are neglected. The physical reason for this is as
follows. In the absence of Brownian rotation, an elongated cell ($\eta
\gg 1$) has a lengthy `pause' when it is almost parallel to the
streamlines of the shear flow. The larger the value of $\eta$, the
longer the pause, and the more robust $\vd$ is in the face of shear.
However Brownian rotation `jostles' the cell out of alignment with the
streamlines, puts a limit on the duration of the pause, and
consequently reduces the robustness of $\vd$ in the face of shear.

\section{Discussion and Conclusion}
\label{sec:conclusion}

In this paper we have used a combined analytic and numeric approach to
calculate the drift velocity $\vd$ of a cell performing run and tumble
chemotaxis in a uniform shear flow, taking account of (i) temporal
comparisons performed by the cell, (ii) persistence of direction, (iii)
Brownian rotation, and (iv) cell geometry. Our key findings are that
(a) shear reduces the drift velocity, and a sufficiently strong shear
may cause the cell to perform negative chemotaxis and swim down the
chemoattractant gradient rather than up it, (b) in terms of maximising
drift velocity, a high persistence parameter is advantageous in a
quiescent fluid but disadvantageous in a shear flow, and (c) a more
elongated body shape is advantageous in performing chemotaxis in a
strong shear flow.

Temporal comparisons of chemoattractant concentration performed by the
cell constitute a form of time delay in its response. The cell biases
its tumble rate according to the history of chemoattractant
concentration it has experienced, rather than according to the
instantaneous concentration or concentration gradient. Negative
chemotaxis arises from the combination of this time delay with rotation
by the shear flow. For instance, a cell may commence swimming up the
chemoattractant gradient on a particular run, but be rotated by the
shear flow to swim down the gradient. Due to the time-delay in its
response, the cell responds as if it were still swimming up the
gradient, so it extends the length of the run, even though the net
displacement is down the gradient. Conversely, a run with a net
displacement up the chemoattractant gradient may be foreshortened.
Lengthened runs down the gradient and foreshortened runs up the
gradient lead to negative chemotaxis. Note that negative chemotaxis
does not occur for an organism that responds to the instantaneous
chemoattractant gradient; the drift velocity given by
(\ref{eq:vdgradientdetection}) is non-negative.

The possibility of negative chemotaxis was already derived by
\citet{Bearon.2000}, although the authors did not comment on it
explicitly. They calculated the drift velocity for a spherical cell
performing run and tumble chemotaxis in a shear flow, and investigated
a form of time delay in the cell's response. In their model, a cell
could bias its tumble rate according to either the present angle
between the swimming direction and the chemoattractant gradient (no
delay), or the original angle between the swimming direction and the
chemoattractant gradient at the beginning of the run (delay). The
authors found that in the delay case the drift velocity was
significantly more reduced by shear than in the no-delay case, and
equation 27 of their paper shows that in the delay case the drift
velocity becomes negative for sufficiently large shear.

The present study shows that in terms of maximising drift velocity, a
high persistence parameter is advantageous in a quiescent fluid but
disadvantageous in a sufficiently strong shear flow.
\citet{Locsei.2007} previously showed that there is a synergy between
temporal comparisons and persistence, and that when temporal
comparisons are taken into account the drift velocity in a quiescent
fluid is maximised by a persistence parameter $\ap \approx 0.78$, which
is greater than the observed value of $\ap \approx 0.33$. In that
paper, it was suggested that the discrepancy between the observed and
apparently optimal values of $\ap$ was due to a compromise between
transient and steady-state behaviour, with lower values of $\ap$
favouring steady state behaviour. The results of the present study
suggest that the observed persistence $\ap \approx 0.33$ may reflect an
additional trade-off between performance in a quiescent fluid and
performance in a sheared fluid.

The present study also shows that the elongated shape of an \emph{E.
coli} cell works to its advantage in terms of performing chemotaxis in
a shear flow, particularly if the rotational diffusivity is zero or
small. A more elongated body has a longer Jeffery orbit period in a
shear flow, and this reduces the likelihood that the shear flow rotates
it from facing up the gradient to down the gradient or vice-versa in
any given run. Thus, a more elongated body maintains a positive drift
velocity up a larger value of shear strength $\Omega$ than a less
elongated body. This effect is still present, but less pronounced, when
rotational diffusivity is accounted for, because the rotational
diffusivity perturbs the Jeffery orbits.

It is worth mentioning that the influence of bacterial shape and
persistence of direction on chemotaxis in a moving fluid has been
investigated in a slightly different context by
\citet{Luchsinger.1999}. In their oceanographically-motivated paper
they numerically modelled bacteria clustering around a localised
nutrient source in a straining flow. They showed that the tendency of
an elongated bacterial cell to align with the direction of the
straining flow assists it to remain near the nutrient source.
Furthermore, they showed that a `back and forth' motion ($\ap = -1$) is
superior to a run and tumble motion ($\ap = 0$) in this context,
because it allows the cell to maintain its alignment with the straining
flow. Their work in conjunction with the present study illustrates that
the `optimal' values for chemotaxis parameters depend on context; a
strategy that works well for locating a localised nutrient source may
be different from a strategy that works well for swimming up a uniform
nutrient gradient.

In conclusion, we should note that while the present work aimed to
improve the realism of previous studies by including effects that were
previously omitted, the model is still somewhat idealised and this
should be born in mind when interpreting the results. More work, both
theoretical and experimental, needs to be done to understand bacterial
chemotaxis in realistic flow conditions. For example, one of the
original motivations of this research was to derive the drift velocity
and effective self-diffusivity for populations of bacteria in a general
linear flow (\ref{eq:linflow}), not just a simple shear directed
perpendicular to the local chemoattractant gradient
\citep{Bearon.2001}. This remains a future goal. In addition,
experiments indicate that in certain conditions boundary effects play
an important role and even allow bacteria to swim upstream
\citep{Hill.2007}, and the fluid dynamics of this are not fully
understood. Finally, we note that one of the strongest motivations for
studying bacterial chemotaxis in moving fluids is to understand the
role of marine bacteria in the pelagic food web, and yet while it is
known that marine bacteria are able to swim faster and react more
rapidly to chemical stimuli than \emph{E. coli}
\citep{Mitchell.1996,Barbara.2003}, their chemotactic response has yet
to be thoroughly characterised and this must be done before modelling
can proceed on a firm footing.


The authors are grateful for enlightening discussions with John
Rallison, Ray Goldstein, Howard Berg, Henry Fu, Roman Stocker, Marcos,
and Rachel Bearon.

\bibliographystyle{spbasic}
\bibliography{everything}


\end{document}